\documentclass[runningheads]{svmult}

\usepackage{makeidx}   
\usepackage{graphicx}  
\usepackage{multicol}  
\usepackage{physprbb}  
\makeindex             

\begin{document}
\title*{Nonfactorizable effects in $B \rightarrow J/\psi K$}
%
%
%
%
%
\author{Bla\v zenka Meli\'c\inst{1,2}
\thanks{Talk given by B.M. at the Workshop on B physics and CP violation, Frontier Science 2002, 06-11 October 2002,
INFN, Frascati, Italy, to appear in the Frascati Physics Series.}
\thanks{Alexander von Humboldt fellow. On leave of absence from
Rudjer Bo\v skovi\'c Institute, Zagreb, Croatia.}
and Reinhold R\"uckl\inst{1}}
%
%
%
\institute{Institut f\"ur Theoretische Physik und Astrophysik, Universit\"at W\"urzburg,
D-97074 W\"urzburg, Germany
\and Institut f\"ur Physik, Universit\"at Mainz, D-55099 Mainz, Germany}

\maketitle              

\begin{abstract}
The method of QCD light-cone sum rules is used to calculate soft nonfactorizable
contributions to the decay amplitude for $B \rightarrow J/\psi K$.
The result confirms expectations that in color-suppressed decays nonfactorizable
corrections can be  sizable.
\end{abstract}

\section{Decay amplitude}
Precise measurements of exclusive nonleptonic $B$ decays have initiated theoretical
considerations which go beyond {\it naive factorization} 
frequently used in the calculation
of decay amplitudes. Nonfactorizable contributions have been investigated
in several approaches \cite{BBNS,KLS,Khodja} for different classes
of two-body nonleptonic $B$ decays.
Here, we focus on 
nonfactorizable corrections in the decay 
$B \rightarrow J/\psi K$. This channel is particularly
interesting because of a substantial discrepancy between the prediction 
from naive factorization and experiment 
by more than a factor of
three in the branching ratio. Also, this mode belongs to the color-suppressed class-2 decays 
for which large nonfactorizable contributions are expected.

The weak matrix element can be written in the form
\begin{equation}
\langle J/\psi K | H_W | B \rangle = \sqrt{2}\, G_F \, V_{c b} V_{c s}^* \, \epsilon \cdot q \,
m_{J/\psi} f_{J/\psi} F_{BK}^+(m_{J/\psi}^2)\, a_2 \, ,
\label{eq:result}
\end{equation}
where $F_{BK}^+$ is the $B \rightarrow K$ form factor and 
the parameter $a_2$ incorporates factorizable and nonfactorizable contributions. A particular  
useful parametrization is given by
\begin{equation}
a_2 = C_2(\mu) + \frac{C_1(\mu)}{N_c} +
2 \, C_1(\mu) \, \frac{\tilde{F}_{BK}^+(\mu)}{F_{BK}^+(m_{J/\psi}^2)} \, ,
\label{eq:a2def}
\end{equation}
where $C_{1,2}$ are the short-distance Wilson coefficients. The first two terms in (2) result from 
naive factorization, while the 
term  proportional to $\tilde{F}_{BK}^+$ represents nonfactorizable contributions. 
Since $a_2$ parametrizes a physical decay amplitude, the scale dependence of the individual terms 
in $a_2$ must cancel.  Taking $\mu = m_b$, one has, numerically,  
$C_1(m_b) = 1.802$ and 
$C_2(m_b) = -0.185$ \cite{BBL}.
%
\section{Nonfactorizable effects}
According to the large $N_c$ analysis \cite{BGR}, there could be a cancellation 
between the last two terms in (2), both being of order $1/N_c$ relative to the 
leading term. In case of such a cancellation, $a_2 \simeq C_2$ would be negative. 
While for $D$ decays a negative value of $a_2$ is consistent with experiment, 
experimental data on $B$ decays suggest a positive 
$a_2$. On the other hand, explicit estimates of nonfactorizable contributions in $B$ decays
have predicted both signs for $a_2$. 

\begin{figure}[t]
\vspace{4.0cm}
\includegraphics{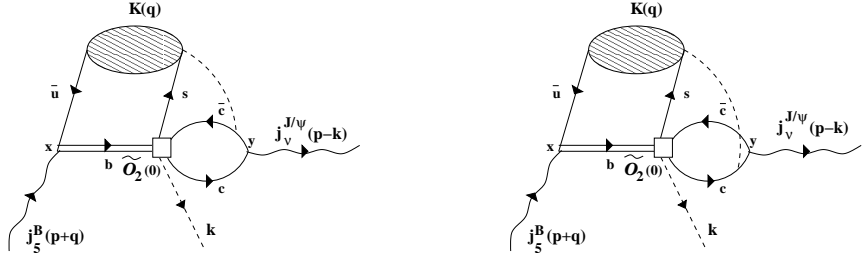}
\caption{\it Leading soft nonfactorizable contributions as estimated by QCD light-cone sum rules. 
The shaded ellipse denotes 
the $K$ meson light-cone distribution amplitude. 
The currents $j^{J/\psi}_{\nu}$ and $j^B_5$ generate states with 
$J/\psi$ and $B$ quantum numbers, respectively. 
The square stands for the ${\tilde {\cal O}}_2$ four-quark
weak operator \cite{MR}. 
\label{deffig} }
\end{figure}

In \cite{MR}, the {\it QCD light-cone sum rule method} \cite{Khodja} was applied to 
the calculation of soft nonfactorizable contributions in $B \rightarrow J/\psi K$.  
The relevant operators in the weak Hamiltonian $H_W$ are 
\begin{equation}
{\cal O}_2 = (\overline{c}\Gamma_{\mu}c)(\overline{s}\Gamma^{\mu}b)\, , \qquad 
\tilde{\cal O}_2 = (\overline{c}\Gamma_{\mu}\frac{\lambda^a}{2}c)(\overline{s}\Gamma^{\mu}
\frac{\lambda^a}{2}b) \, .
\end{equation}
It is the contribution of $\tilde{\cal O}_2$ to the matrix element (1),  which is expected 
to give rise to the leading nonfactorizable effects. The sum rule approach allows to 
estimate the contribution of soft-gluon exchange 
between the $J/\psi$ and the $B-K$ system, 
see Fig. 1. 
To this end, one has to isolate the ground state contribution to the correlation 
function 
\begin{equation}
F_{\nu}(p,q,k)= \int d^4 x e^{-i(p+q)x} \int d^4 y e^{i(p-k)y} \langle K(q) |
T \{ j_{\nu}^{J/\psi}(y) \tilde{\cal O}_2(0) j_5^{B}(x) \} | 0 \rangle
\label{eq:corr0}
\end{equation}
represented graphically in Fig.1. Taking into account 
twist-3 and twist-4 contributions calculated at the appropriate scale 
$\mu_b \sim m_b/2$, one finds
\begin{equation}
\tilde{F}^+_{BK}(\mu_b) = 0.009 \div 0.017\, ,
\label{eq:Ftilde}
\end{equation}
and, substituting (5) in (2), 
\begin{equation}
a_2 = 0.14 \div 0.17\,|_{\mu = \mu_b}\, .
\end{equation}
Although the nonfactorizable matrix element (5) is rather small, it 
increases $a_2$ by $30 \div 70 \%$, due to the large coefficient $2 C_1(\mu_b)$. 

In addition, one has nonfactorizable contributions from hard-gluon exchange. 
They 
have been 
estimated in {\it QCD factorization} \cite{BBNS} and amount 
to another $20 \%$ correction \cite{Cheng}. Thus, in total one
obtains
\begin{equation}
a_2 = 0.16 \div 0.19\,|_{\mu = \mu_b}\, .
\end{equation}

\section{Conclusions}

Firstly, one can see that the theoretical expectation (7) 
is still too small to explain the experimental value 
\begin{equation}
|a_2|^{exp} = 0.29 \pm 0.03 \, .
\end{equation}
Secondly, it is interesting to note that the theoretical approach described here predicts a 
positive sign of $a_2$ in agreement 
with experiment and in contradiction with the argument based on 
$1/N_c$ expansion \cite{BGR}. 
Finally, a comparison of the value (8) deduced from the measurement of $B \rightarrow J/\psi K$ 
with the result $|a_2|^{exp} = 0.4 \div 0.5$ from the decays $\overline{B}^0 \rightarrow D^{(\ast)0}\pi^0$ 
indicates a substantial nonuniversality of $a_2$ in color-suppressed decays \cite{NP}.


%
%
%
\section*{Acknowledgments}
B.M. wishs to thank the organizers of Frontier Science 2002 for the invitation and for their support 
and would also like to acknowledge the support by the Alexander
von Humboldt Foundation and partial support of the Ministry of Science and Technology of the 
Republic of Croatia under the contract 0098002. R.R. acknowledges the support by the 
Bundesministerium f\"ur Bildung und Forschung (BMBF, Bonn, Germany) under the contract number 05HT1WWA2.

\end{document}